# Experimental search for neutron – mirror neutron oscillations using storage of ultracold neutrons

A.P. Serebrov [a†], E.B. Aleksandrov [b], N.A. Dovator [b], S.P. Dmitriev [b],
A.K. Fomin [a], P. Geltenbort [c], A.G. Kharitonov [a], I.A. Krasnoschekova [a],
M.S. Lasakov [a], A.N. Murashkin [a], G.E. Shmelev [a], V.E. Varlamov [a],
A.V. Vassiljev [a], O.M. Zherebtsov [a], O. Zimmer [c,d]

[a] *Petersburg Nuclear Physics Institute, RAS, 188300, Gatchina, Leningrad District, Russia*

[b] *Ioffe Physico-Technical Institute, RAS, 194021, St. Petersburg, Russia*

[c] *Institut Laue-Langevin, BP 156, 38042 Grenoble cedex 9, France*

[d] *Physik-Department E18, TU München, 85748 Garching, Germany*

**Abstract**

The idea of a hidden sector of mirror partners of elementary particles has attracted considerable interest as a possible candidate for dark matter. Recently it was pointed out by Berezhiani and Bento that the present experimental data cannot exclude the possibility of a rapid oscillation of the neutron n to a mirror neutron n′ with oscillation time much smaller than the neutron lifetime. A dedicated search for vacuum transitions n → n′ has to be performed at weak magnetic field, where both states are degenerate. We report the result of our experiment, which compares rates of ultracold neutrons after storage at a weak magnetic field well below 20 nT and at a magnetic field strong enough to suppress the seeked transitions. We obtain a new limit for the oscillation time of n-n' transitions, $\tau_{osc}$ (90% C.L.) > 414 s. The corresponding limit for the mixing energy of the normal and mirror neutron states is $\delta m$ (90% C.L.) < 1.5×10$^{-18}$ eV.

[†] corresponding author, tel. +7-81371 46 001, e-mail: serebrov@pnpi.spb.ru

# 1 Introduction

As suggested long ago there might exist a parallel sector of "mirror particles" in form of a hidden duplicate of the observable particle sector [1]. It would provide a natural explanation of parity violation in the ordinary weak interactions in terms of an exact discrete symmetry, the so-called mirror parity, with respect to interchange of particles between the ordinary and a mirror world. In this picture, ordinary and mirror sectors should have identical particle contents and identical microphysics, in particular same mass spectra and coupling constants (gauge, Yukawa, Higgs). Each ordinary particle, i.e. electron, nucleon etc. is supposed to have its mirror twin exactly degenerate in mass but being sterile with respect to the ordinary gauge interactions. Vice versa, ordinary particles should be sterile with respect to the mirror gauge forces. The gravitational interactions are supposed to exist between the two sectors. Nowadays this hypothesis becomes increasingly popular for its intriguing implications in particle physics and cosmology (for recent reviews, see [2]). Mirror matter could be a viable dark matter candidate, with a variety of interesting applications for key cosmological issues such as inflation, baryogenesis, nucleosynthesis, recombination, formation of cosmological structures, gravitational lensing, etc., which can be tested by cosmological observations [3].

Besides gravity, the two sectors could communicate also by other means. In particular, no fundamental conservation law prohibits any neutral ordinary particle, elementary or composite, to mix with its mirror partner. The "ordinary-mirror" oscillation phenomenon thus possible would render the search for mirror matter amenable to terrestrial experiments. For instance, the kinetic mixing between the ordinary and mirror photons [4] can be investigated searching for a transition of positronium to mirror positronium [5]. The mass mixing between the ordinary and mirror neutrinos could be revealed in the active-sterile neutrino oscillations [6]. Also ordinary pions and Kaons could have a mass mixing with their mirror partners which can be induced e.g. by some extra gauge forces between the elementary particles of the two sectors [7].

From the phenomenological viewpoint, the small mass mixing between the ordinary neutron n and its mirror partner n' leads to the most intriguing possibility. As it was shown in ref. [8], the existing experimental limits on n-n' oscillations are very weak, allowing the oscillation time $\tau = 1/\delta m$ to be much smaller than the neutron decay time $\tau_n \sim 10^3$ s. This could have direct astrophysical consequences, in particular, for the propagation of ultra-high energy cosmic rays at cosmological distances [8] or of the neutrons from the solar flares [9]. The experimental possibilities to test the n-n' oscillation were discussed in detail in ref. [10].

From the theoretical viewpoint, such a mixing can emerge from the effective six-fermion interactions $G(udd)(u'd'd')$ with a dimensional coupling constant $G = 1/M^5$ where $M$ is the relevant mass scale and u,d and u'd' are the ordinary and mirror quarks, leading to $\delta m \sim (10 \text{ TeV}/M)^5 \cdot 10^{-15}$ eV. Such effective interactions can be induced via exchange of extra heavy states mediating between the ordinary and mirror sectors for which corresponding renormalizable models were constructed [8,9].

Since mirror neutrons are invisible, the n-n' oscillation can manifest experimentally only as a neutron disappearance. However, as shown in ref. [8], the neutron cannot disappear from a stable nucleus and thus does not induce nuclear instabilities, which is in a drastic difference to the case of neutron-antineutron oscillations [11]. On the other hand, the n-n' oscillation is very sensitive to external conditions: it is suppressed by matter effects and, remarkably, by the magnetic field of the earth. Clearly, the possibility of a fast oscillation process which violates baryon number conservation looks rather intriguing (for comparison, the direct experimental limit on the neutron - antineutron oscillation time is $8.6\times10^7$ s [12] while the limit obtained from nuclear stability is about the same, $1.3\times10^8$ s [13]).

## 1 Method to search for n-n' transitions

The evolution of a free neutron to the mirror state is described by the effective Hamiltonian [8]

$$H = \begin{pmatrix} \frac{p^2}{2m} - i\frac{\Gamma}{2} + V & \delta m \\ \delta m & \frac{p^2}{2m} - i\frac{\Gamma}{2} + V' \end{pmatrix}. \qquad (1)$$

The mass $m$ and the decay width $\Gamma = 1/\tau_n$ are precisely the same for ordinary and mirror neutrons due to exact mirror parity, but the potential energies $V$ and $V'$ of n and n' are in general different. The energy $V = -\mu_n H$ of the neutron magnetic moment $\mu_n \approx 6\times10^{-8}$ eV/T in an ordinary magnetic field $H$ is absent for the mirror neutron. Since the presence of significant mirror magnetic fields on the earth seems extremely unlikely, $\Delta V = V - V' = -\mu_n H$.

The probability of transition from the initial neutron state at time 0 to a mirror neutron state at time $t_f$ of free neutron flight is given by the following equation [8,10]:

$$P_{nn'}(t_f) = \frac{\delta m^2}{\delta m^2 + \omega^2} \cdot \sin^2\left(t_f \sqrt{\delta m^2 + \omega^2}\right), \qquad (2)$$

where $\omega = \mu_n H/2$ (here and in the following natural units are used, $c = \hbar = 1$). The factor $\exp(-\Gamma t_f)$ can be neglected in eq. (2) once we consider time intervals much shorter than the lifetime of the neutron, $t_f \ll \tau_n$. When the magnetic field is so weak that $\omega \ll \delta m$, a full conversion of neutrons to mirror neutrons is possible, i.e. the transition probability $P_{nn'}(t_f)$ may reach unity. Equation (2) then simplifies to

$$P_{nn'}(t_f) = \sin^2\left(t_f/\tau_{osc}\right), \qquad (3)$$

where $\tau_{osc} = 1/\delta m$ is the oscillation time. If on the other hand the magnetic field is strong enough, $\omega \gg \delta m$, eq. (2) becomes

$$P_{nn'}(t_f) = \frac{\sin^2 \omega t_f}{(\omega \tau_{osc})^2}, \qquad (4)$$

In this case the period of n-n' oscillations will be a factor 4 larger than the period of neutron Larmor precession. The maximum probability of transition to the mirror neutron is suppressed by the factor $1/(\omega\tau_{osc})^2$. Figure 1 illustrates this for different values of $\omega\tau_{osc}$. The transition probability is not suppressed if $\omega t_f \ll 1$ but it may be strongly suppressed if $\omega t_f \gg 1$.

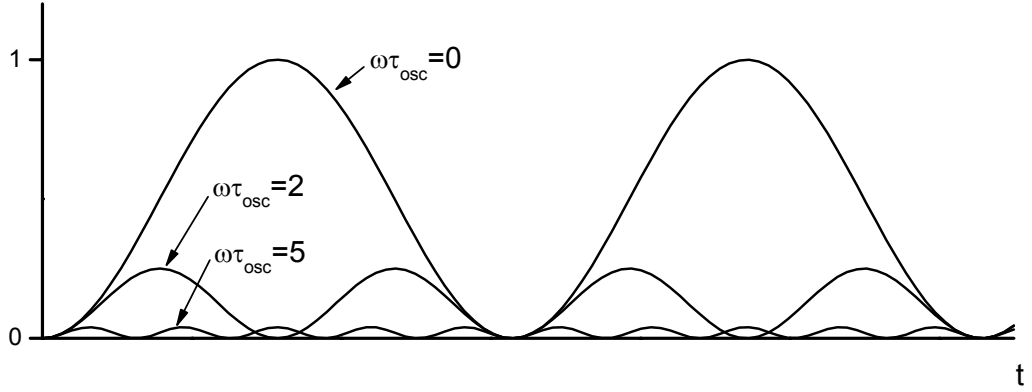

**Figure 1.** The probability of n-n' oscillations as function of time for different values of $\omega\tau_{osc}$.

Experiments are restricted to short times $t_f$. Using cold neutrons with velocity $v \approx 1000$ m/s, on a path of 100 m it would be only $t_f = 0.1$ s. The same value can be obtained in an experiment with ultracold neutrons (UCN, $v \approx 5$ m/s) when a trap is employed with mean free path 0.5 m. With the trap well shielded from the earth's magnetic field, $\omega t_f \ll 1$, eq. (4) turns to the form analoguous to eq. (3),

$$P_{nn'}(t_f) = \sin^2(t_f/\tau_{osc}). \qquad (5)$$

The transition probability is not significantly affected by a magnetic field of 10 nT, when the time of free flight of the neutron is 0.1 s. In the opposite case, $\omega t_f \gg 1$, the average of the oscillating term in eq. (4) is $1/2$ thus leading to

$$P_{nn'}(t_f) = \frac{1}{2(\omega\tau_{osc})^2}. \qquad (6)$$

In order to attain a suppression factor 10, already a magnetic field of 0.5 μT will be sufficient.

The main idea of the experiment is to measure the difference of storage times of UCN in a trap with magnetic field switched on and off. When switched off ("zero"), the transition of neutron to mirror neutron can happen during the free flight of the neutron between wall collisions. If it takes place the mirror neutron will pass through the trap wall without interaction. As a result the storage time will be reduced. When the magnetic field is switched on, n-n' transitions are suppressed, and the storage time is defined by the probability of neutron β-decay and the probability of losses in the reflection from the trap walls. In our experiment the residual magnetic field inside a four-layer permalloy shielding reached values down to 2 nT, corresponding to $\omega t_\text{f} = 10^{-2}$. Measurements with field were performed with 2 µT, leading to a suppression factor $2(\omega t_\text{f})^2$ of $1.7 \times 10^4$.

Degeneracy of the neutron and mirror neutron states requires, besides a low magnetic field, a good vacuum inside the trap for UCN. A pressure of $1 \times 10^{-5}$ mbar corresponds to a potential of neutron interaction with the residual gas of $0.5 \times 10^{-15}$ eV. The corresponding value $\omega_\text{vac} t_\text{f} = 0.05$ is enough to carry out the experimental search for n-n' transitions.

Using the equations presented above it is easy to obtain the relation between the difference of UCN storage times with the magnetic field switched on and switched off, and the value $\tau_\text{osc}$. The probability of UCN losses per second in the trap for vanishing magnetic field is given by

$$\tau_{\text{st},0}^{-1} = \tau_\text{n}^{-1} + \mu \nu + \frac{\langle t_\text{f}^2 \rangle}{\tau_\text{osc}^2} \nu, \quad (7)$$

where $\tau_{\text{st},0}$ is the corresponding storage time of UCN, $\mu$ is the loss probability per single collision with trap walls, $\nu$ is the average frequency of collision, $\langle t_\text{f}^2 \rangle$ is the mean square time of neutron's free flight in the trap. When magnetic field $H$ is switched on the loss probability is given by

$$\tau_{\text{st},H}^{-1} = \tau_\text{n}^{-1} + \mu \nu, \quad (8)$$

and the difference by

$$\tau_{\text{st},0}^{-1} - \tau_{\text{st},H}^{-1} = \frac{\langle t_\text{f}^2 \rangle}{\tau_\text{osc}^2} \nu. \quad (9)$$

The determination of storage time requires counting of the number $N_1$ of UCN in the trap after a short holding time $t_1$ (for normalization of initial number of neutrons in the trap) and $N_2$ after a longer holding time $t_2$,

$$\tau_\text{st} = (t_2 - t_1) / \ln(N_1 / N_2). \quad (10)$$

For optimal strategy of measurements with respect to counting statistics, $t_2$ is chosen about $2\tau_\text{st}$.

A first method to determine $\tau_\text{osc}$ consists in measuring the difference $\theta = \tau_{\text{st},H} - \tau_{\text{st},0}$. From eq. (9) we obtain

$$\tau_\text{osc}^{-2} = \frac{\theta}{\langle t_\text{f}^2 \rangle \nu} \cdot \tau_\text{st}^{-2}. \quad (11)$$

If no difference between $\tau_{\text{st},H}$ and $\tau_{\text{st},0}$ is found, one can establish a limit for $\tau_\text{osc}$, assuming Gaussian errors. With confidence level 90% it is given by

$$\tau_\text{osc}^{\text{C.L. 90\%}} \geq \tau_{\text{st},0} \sqrt{\langle t_\text{f}^2 \rangle \nu / (\theta + 1.65 \Delta \theta)} \qquad \theta > 0$$

$$\tau_\text{osc}^{\text{C.L. 90\%}} \geq \tau_{\text{st},0} \sqrt{\langle t_\text{f}^2 \rangle \nu / 1.65 \Delta \theta} \qquad \theta < 0, \quad (12)$$

where $\Delta\theta$ is the experimental uncertainty of $\theta$.

In a second method to determine $\tau_{osc}$ one measures the ratio $R$ of the registered neutrons $N_0$ and $N_H$ without and with magnetic field after the holding time $t_h$,

$$R \equiv N_0 / N_H = e^{-\langle t_f^2 \rangle v (t_h + \tau_{fill} + \tau_{emp})/\tau_{osc}^2} = 1 - \langle t_f^2 \rangle v (t_h + \tau_{fill} + \tau_{emp})/\tau_{osc}^2. \quad (13)$$

Since the n-n' transistions may also happen during filling and emptying the trap, the effective holding time $t_h + \tau_{fill} + \tau_{emp}$ enters instead of $t_h$. Defining $r = 1 - R$ the oscillation time is given by

$$\tau_{osc}^{-2} = \frac{r}{\langle t_f^2 \rangle v (t_h + \tau_{fill} + \tau_{emp})}. \quad (14)$$

This way to determine $\tau_{osc}$ is more sensitive than via the $\theta$-measurement, also because measurements both at $t_1$ and $t_2$ can be used to calculate ratios $R$, identifying times $t_{h1} \equiv t_1$ and $t_{h2} \equiv t_2$.

Figure 2 illustrates the difference of the two methods. In the $\theta$-measurement we search for the change of storage time, in the $r$-measurement we search for the change of the ratio $R$. In principle the measurement by means of R-ratio is enough to install limit for n-n' transition. But in case if there is some positive signal the confirmation by means of θ-measurement is obligatory to demonstrate that effect is connected with changing of storage time and is not connected with changing of number of trapped UCN for different magnetic field. For higher reliability of the final result we decided to perform most measurements with two different holding times in order to be able to analyse the data using both methods.

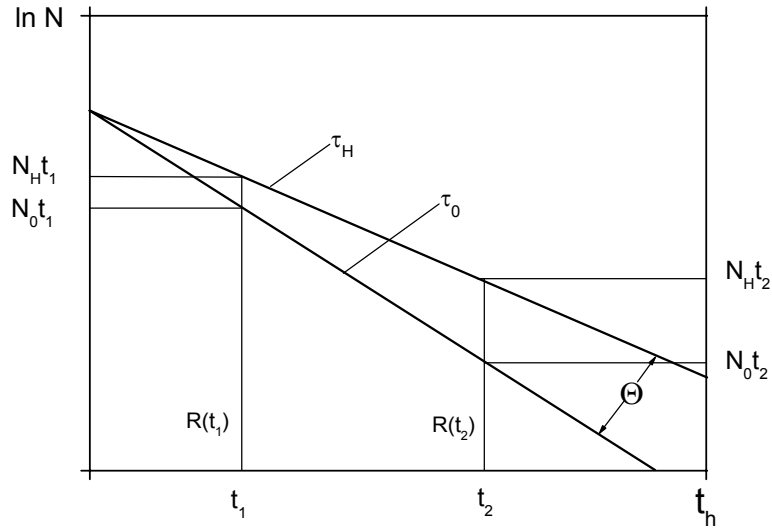

**Figure 2.** Illustration of the difference of two methods: $\theta$-measurement and $r$-measurement. $N$ is the number of neutrons in the trap for different holding times.

It should be mentioned that the realization of an experiment with storage of UCN seems much easier than a beam experiment. The installation is compact, and does not require a voluminous magnetic shielding on a long flight base. Also stability of neutron beam intensity is not as severe an issue as in a beam experiment. In UCN storage each neutron is used a few thousand times and the effect is proportional to $n(t_f/\tau_{osc})^2$ where $n$ is the number of free flights between wall collisions during the holding time in the trap. A simple analysis shows that for storage times less than the neutron lifetime the sensitivity of experiment is proportional $L^{3/2}\mu^{-1/4}\rho^{1/4}T^{1/4}$, where $L$ is the representative linear size of trap, $\mu$ is the loss factor per collision, $\rho$ is the UCN density, and $T$ is the total time of measurements. Thus the decisive role plays the size of trap. It was therefore chosen as large as possible.

## 2  Experimental installation to search for n-n' transitions

The experiment has been carried out using the well-known UCN facility PF2 of the ILL reactor. It employs the vacuum chamber of the new PNPI spectrometer to search for the electric dipole moment of the neutron [14], which for this purpose was adapted to the search for n-n' transitions. The scheme of installation is shown in Fig. 3.

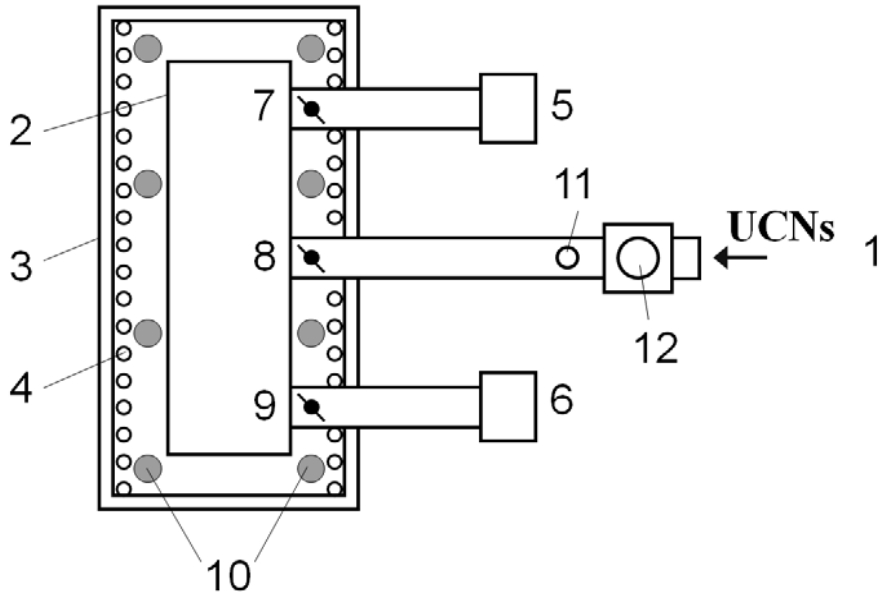

**Figure 3.** Experimental setup (top view). 1: UCN input guide; 2: UCN storage chamber; 3: magnetic shielding; 4: solenoid; 5-6: UCN detectors; 7-9: valves; 10: Cs-magnetometers, 11: monitor detector, 12: entrance valve.

The whole arrangement consists of a UCN storage trap with valves for filling and empting of UCN, a neutron guide system, UCN detectors and a magnetic shielding. UCN enter the trap (2) through the input UCN guide (1), with the valves (7) and (9) closed. After filling the trap during 100 s, the valve (8) is closed and UCN are kept in the trap during the given holding time $t_1$ or $t_2$. Then the valves (7) and (9) are opened and neutrons reach UCN detectors (5 and 6). Count rates recorded by these detectors during the whole process are shown in Fig. 4. The two UCN detectors offer the possibility for cross checking the detector stability. In the data analysis the summed counts of both detectors were used.

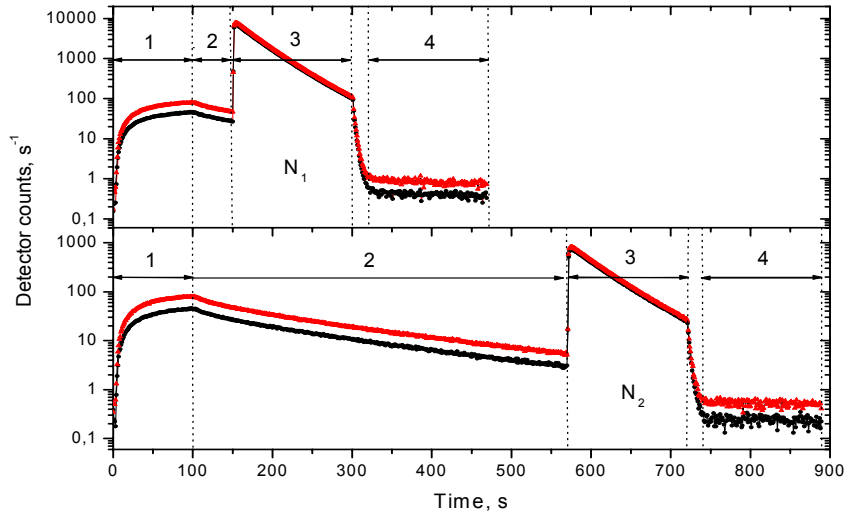

**Figure 4.** Count rates of UCN detectors (5 and 6) in log scale during measurements. The filling time is 100 s. Holding times were $t_1 = 50$ s and $t_2 = 470$ s. Empting time is 150 s. The time of background measurement is 150 s. The region 3 in these plots is used to deduce the numbers $N_1$ and $N_2$ required for the determination of the storage time, respectively the ratio $R$ (after background subtraction). This picture was obtained after 130 cycles. $\tau_{\text{fill}} = 35$ s, $\tau_{\text{emp}} = 30$ s after holding time 50 s, $\tau_{\text{emp}} = 38$ s after holding time 470 s.

The whole process (filling, holding, emptying and background measurement) is controlled by means of the UCN detectors (5 and 6). Although the valves (7 and 9) are closed during filling and holding times, UCN counts are recorded due to small slits. The effective slit area of the closed valves was about 0.2 mm. This small aperture does not significantly deteriorate the storage time but allows to monitor the processes of filling and storage.

The stability of the incident UCN beam during filling was measured by means of a monitor detector (11), to which the other UCN count rates were normalized, i.e. instead of the ratio $N_1/N_2$ the ratio $N_1 M_2/N_2 M_1$ was used in eq. (10), where $M_1$ and $M_2$ are the count rates of the monitor detector during the two filling times. Similar corrections were used in eq. (13).

The UCN trap with volume 190 liters is a horizontal cylinder with diameter 45 cm and length 120 cm. It was made from copper with the inner surface coated by beryllium. The critical velocity of this coating is 6.8 m/s.

Using Monte-Carlo calculation we estimated the average frequency of collisions of UCN with trap walls, $v$, as well as $\langle t_f^2 \rangle$. These values are slightly different for different holding times. For example $v = 11.11$ s$^{-1}$ and $\langle t_f^2 \rangle = 0.011$ s$^2$ for holding time 50 s, $v = 10.43$ s$^{-1}$ and $\langle t_f^2 \rangle = 0.013$ s$^2$ for holding time 470 s. The parameter $\langle t_f^2 \rangle v$ is equal to 0.126 s for the holding time 50 s and 0.137 s for the holding time 470 s.

The UCN storage times were deduced from measurements with different holding times using eq. (10). We obtained 208 s for holding times $t_1$ = 50 s and $t_2$ = 470 s, 198 s for holding times $t_1$ = 50 s and $t_2$ = 350 s, and 188 s for holding times $t_1$ = 50 s and $t_2$ = 250 s. The variation is due to the spectral dependence of the storage time.

The magnetic shielding of the installation consists of four layers of permalloy. The residual magnetic field inside the shielding was about 2 nT, which is low enough to carry out the search for n-n' transitions. The level of residual field was controlled by means of 14 Cs-magnetometers which were installed around of UCN trap [15]. For monitoring the transition-suppressing magnetic field 2 µT, two additional Cs-magnetometers were used.

## 3 Results of measurements

Measurements were carried out in a mode, which allows to remove the drift of incident UCN beam intensity and a possible drift of the UCN storage time. For this purpose the measurements with different holding times $t_1$ and $t_2$ were interchanged in the sequence ($t_1$, $t_2$, $t_2$, $t_1$). Such a sequence provides one measurement of $\tau_{st}$ at the "zero" magnetic field or at the "suppressing" magnetic field $H$. The measurements with magnetic field "0" and "$H$" were interchanged in sequence (0, $H$, $H$, 0), ($H$, 0, 0, $H$), ($H$, 0, 0, $H$), (0, $H$, $H$, 0). Such a sequence of measurements allows to remove, besides a linear drift, also an eventual squared long term drift of storage time. A small part of the experiments was carried out in the mode of $r$-measurement, i.e. using only one holding time.

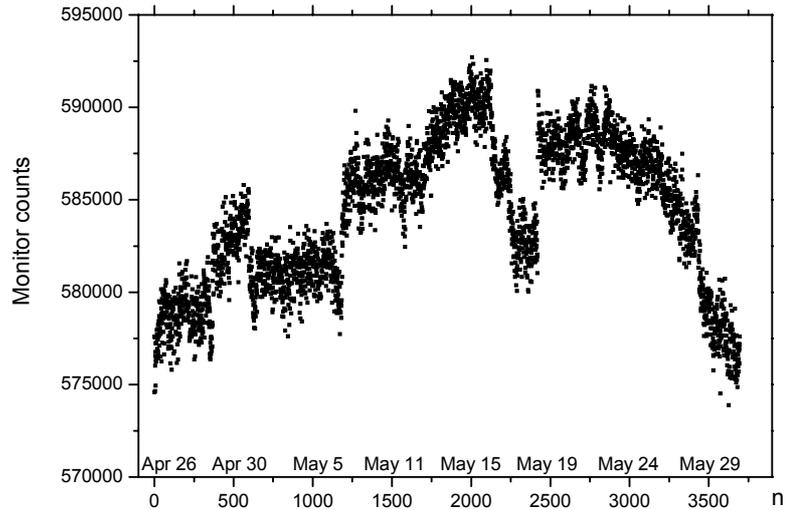

**Figure 5.** Counts recorded with the monitor detector during the measurements. n counts the number of each procedure including UCN filling, UCN keeping and empting.

The total numbers of neutrons registered by both detectors was $5\times10^5$ after holding time 50 s, $3.16\times10^5$ after 150 s, $1.71\times10^5$ after 250 s, $1.1\times10^5$ after 350 s, and $0.65\times10^5$ after 470 s. Correspondingly the accuracy of determination of the $R$-ratio, respectively, the $r$-value ranged between $2\times10^{-3}$ for 50 s and $5.5\times10^{-3}$ for 470 s. The instability of the UCN density determined by the monitor detector during two consecutive measurements of storage time was about $2.3\times10^{-3}$, i.e. it was on the level of the counting statistical accuracy of measurements. The count rate of the monitor detector during the measurements is shown in Fig. 5. Occasionally the changes of UCN intensity reached about $1 - 2$ %, which demonstrates the usefulness of the monitor detector for high reliability.

Figures 6 a and b show the instability of the "zero" magnetic field, as measured with 11 Cs-magnetometers for the same period of time as in Fig. 5. As already stated before the residual magnetic field was about 2 nT. To reach this level required not only the usual demagnetization of the shielding but also electric isolation of the spectrometer from the UCN guide coming from the UCN turbine. This has allowed us to suppress both variable and constant magnetic field components inside the screen, presumably caused by leakage currents between neutron turbine and magnetic screen.

Jumps in the magnetic field visible in Fig. 6 were due to changes of position of a 6-ton reactor crane in only 5 m distance from the spectrometer. In addition, a good fraction of measurements had to be performed with strong magnetic noise produced by a neighbouring installation, where a magnetic coil with effective volume 120 liters and magnetic field 11 mT was switching on and off with time interval of about 1000 s. This coil was placed in only 4 m distance from our spectrometer, resulting in magnetization of our shielding and thus creating a residual magnetic field up to about 15 nT. During all measurements at "zero" magnetic field its actual value did not exceed 20 nT, corresponding to a value of $\omega t_f$ less than 0.1, which is still acceptable to perform the experimental search for neutron – mirror neutron transitions.

As mentioned before, the main part of experiments was carried out to be able to determine $\theta$, which requires measurements at two holding times $t_1$ and $t_2$. Results are presented in Table 1, and the distribution of values of $\theta$ is shown in Fig. 7. Deviations from the average value were normalized to the statistical error, thus enabling us to see the broadening of the distribution caused by other reasons. This could for instance be due to non-reproducibility of the effective slit area of the closed valve. The actual value of the broadening is only 1.1, determined with statistical accuracy $2.5\sigma$, meaning that the dispersion of $\theta$ is still in reasonable agreement with a purely statistical error of the measurements.

a)

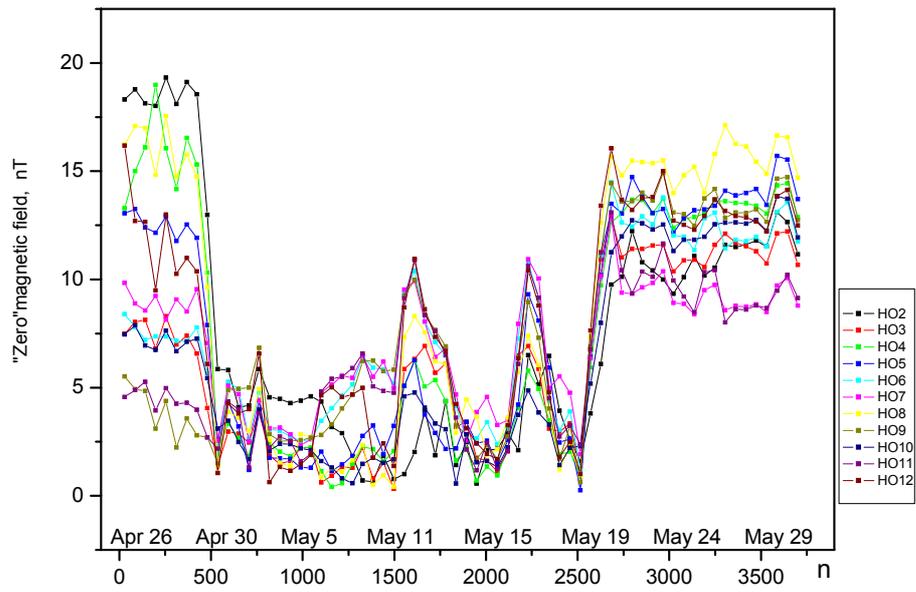

b)

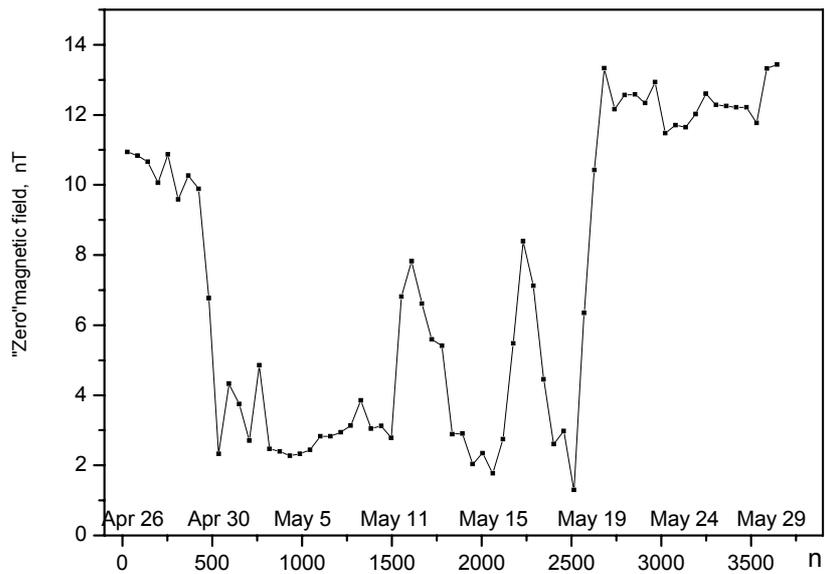

**Figure 6 a,b.** a) Instability of the "zero" magnetic field measure with 11 Cs-magnetometers
b) averaged value.
Shown is the absolute value of the magnetic field component along the axis of spectrometer.

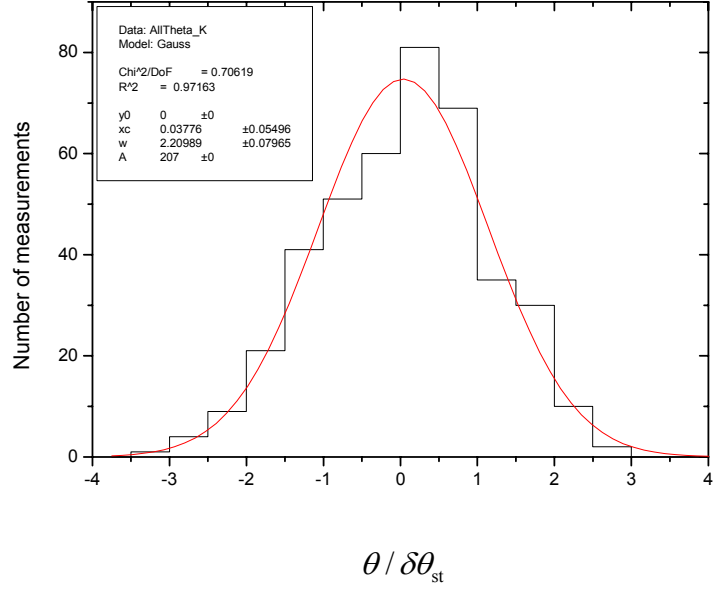

$\theta / \delta\theta_{st}$

**Figure 7.** Histogram of $\theta$-measurements. The solid line shows a Gaussian fit. $\delta\theta_{st}$ is the statistical error of each measurement by means both detectors for the following the sequence ($t_1$, $t_2$, $t_2$, $t_1$) at "zero" magnetic field and at the "suppressing" magnetic field $H$. The width of distribution is $2\sigma = 2.21$ i.e. it is widened with respect to the statistical one by the factor 1.1.

**Table 1.** Result of $\theta$-measurements: $\tau_{osc}^{-2}$ and its distributions are given in units $10^{-6}$ s$^{-2}$. $k$ is the number of each run with about three days length, $\Delta(\tau_{osc}^{-2})_{dis}$ is the uncertainty of measurement deduced from the dispersion of the data of each run, $\delta(\tau_{osc}^{-2})_{st}$ is the counting statistical error of measurement. The total result was calculated with $\Delta(\tau_{osc}^{-2})_{dis}$.

| $k$ | $\tau_{osc}^{-2}$ | $\Delta(\tau_{osc}^{-2})_{dis}$ | $\delta(\tau_{osc}^{-2})_{st}$ | $t_1$ [s] | $t_2$ [s] |
|---|---|---|---|---|---|
| 1 | 12.80 | 10.77 | 11.39 | 50 | 350 |
| 2 | -1.141 | 8.567 | 8.067 | 50 | 470 |
| 3 | 5.778 | 15.60 | 14.48 | 50 | 350 |
| 4 | 21.77 | 11.90 | 11.26 | 50 | 250 |
| 5 | -1.557 | 11.96 | 10.39 | 50 | 350 |
| 6 | -2.677 | 14.54 | 14.54 | 50 | 350 |
| 7 | 3.291 | 11.88 | 11.88 | 50 | 350 |
| 8 | 18.42 | 23.08 | 15.19 | 150 | 470 |
| 9 | 54.62 | 43.94 | 34.70 | 150 | 250 |
| 10 | -5.958 | 11.82 | 11.82 | 150 | 470 |

The final result of 10 runs of $\theta$-measurement is:

$$\tau_{osc,\,\theta}^{-2} = (+7.05 \pm 5.66) \times 10^{-6} \text{ s}^{-2},$$

We can not interpret this result as a positive signal because it differs from zero only by 1.25 standard deviations. Using eq. (12) we therefore state this result as a lower limit on the oscillation time,

$$\tau_{osc,\theta}(90\% \text{ C.L.}) \geq 247 \text{ s}.$$

As stated after eq. (14) the accuracy of measurements is higher if we consider the *r*-values. Results for the normalized *r*-values are shown in Table 2 and in Fig. 8 for the same period of measurements as in Figs. 4 and 5.

**Table 2.** Result of data treatment by *r*-method: Meaning and units of symbols are the same as in Table 1.

| k | $\tau_{osc}^{-2}$ | $\Delta(\tau_{osc}^{-2})_{dis}$ | $\delta(\tau_{osc}^{-2})_{st}$ | $t_1$ [s] | $t_2$ [s] |
|---|---|---|---|---|---|
| 1 | 9.711 | 9.082 | 8.272 | 50 | 350 |
| 2 | -4.95 | 7.396 | 6.734 | 50 | 470 |
| 3 | 10.12 | 9.969 | 8.560 | 50 | 350 |
| 4 | 21.73 | 9.204 | 8.189 | 50 | 250 |
| 5 | 2.147 | 7.908 | 7.583 | 50 | 350 |
| 6 | 3.202 | 10.81 | 10.62 | 50 | 350 |
| 7 | -17.66 | 11.27 | 8.651 | 50 | 350 |
| 8 | -4.006 | 11.82 | 8.386 | 150 | 470 |
| 9 | -12.70 | 9.438 | 8.476 | 150 | 250 |
| 10 | -5.256 | 10.89 | 8.622 | 150 | 470 |
| 11 | 9.090 | 9.514 | 8.234 | $t_h = 350$s | |

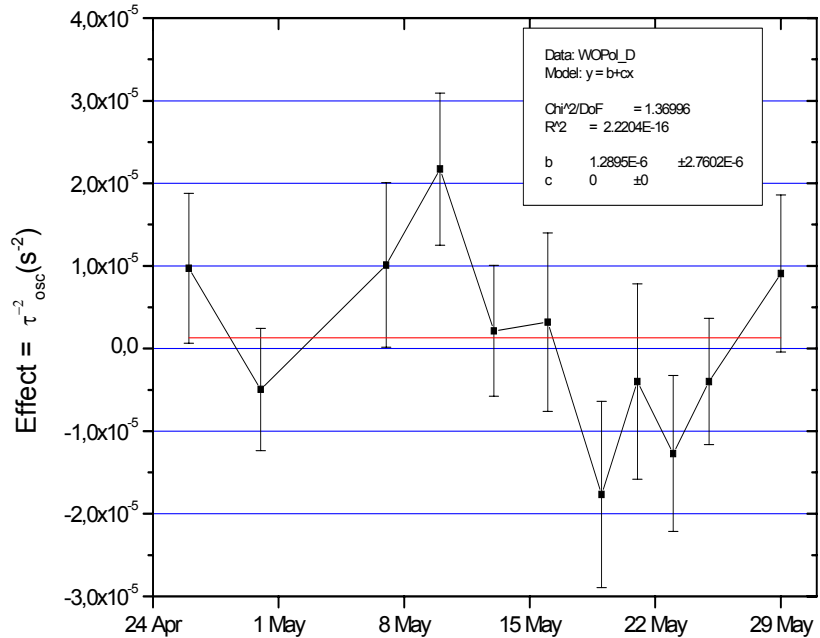

**Figure 8.** The result of measurement of the normalized r-value ($\tau_{osc}^{-2}$). The measurement errors for different runs were calculated from dispersion of measurements inside each run.

A fit of this data by a constant gives the result:

$$\tau_{osc,r}^{-2} = (+1.29 \pm 2.76) \times 10^{-6} \text{ s}^{-2}.$$

The $\chi^2$-value of the data distribution was equal to 1.37 which is acceptable for 11 points. Interpreted as a lower limit on the oscillation time we obtain

$$\tau_{osc,r}(90\% \text{ C.L.}) \geq 414 \text{ s}.$$

This limit is considerably better than limit established in the work [16] $\tau_{osc}$(95% C.L.) ≥ 103 s, in particular if one takes into account that the limit depends on counting statistic as $N^{1/4}$. Since the accuracy of $\theta$-measurements is considerably less than of $r$-measurements, the lower limit for $\tau_{osc}$ established from $r$-measurements covers also the lower limit from θ-measurements, which serves as a control measurement.

## 4  Conclusion

As a result of measurements carried out in this work a new lower limit for the time of neutron mirror neutron oscillations was established:

$$\tau_{osc}(90\% \text{ C.L.}) \geq 414 \text{ s}$$

This limit is already not too far from the neutron lifetime but might still be too low to provide restriction of the mechanism of appearance of high-energy protons above the GZK-cutoff in cosmic radiation due to n-n' oscillations.

Another remark (private communication by B.Kerbikov) concerns the fact that n-n' oscillations are influenced by the fact that the neutron is confined within the trap while the mirror neutron doesn't interact with the trap. The corresponding quantum mechanical problem deserves a dedicated investigation.

In this work it was shown that UCN storage is indeed a very effective experimental method to search for the n-n' transitions. An improvement by a factor 2 may be reached due to increasing the storage volume to a trap diameter of 1 m. Another factor 3 will be available when a UCN density of $10^3$ cm$^{-3}$ will be available from new powerful UCN sources.

More substantial improvement in the search for n-n' transitions might be realized in a new-proposed neutron-antineutron transition search proposed in ref. [17] for DUSEL laboratory. This project assumes the preparation of a vertical path (1 km) with time of flight of cold neutrons about 1s and with a of neutron beam intensity $10^{12}$ s$^{-1}$. The same installation with small modification can be used for the n-n' experiment. By using a very precise monitoring system and an integral method to detect the neutron intensity it might be possible to reach in the n-n' experiment a sensitivity up to $10^4 - 10^5$ s. This would represent an important progress in efforts to find mirror dark matter in laboratory conditions.

**Acknowledgements**

We would like to thank Z. Berezhiani, Yu. Pokotilovski, Yu. Kamyshkov, L.B. Okun, B. Kerbikov and V.A. Kuzmin for very useful and constructive discussions.